\documentstyle[12pt]{article}
\renewcommand{\theequation}{\arabic{section}.\arabic{equation}}
\begin{document}

\title{
       De Rham-Kodaira's Theorem and Dual Gauge Transformations
      }

\author{
        Hisashi ECHIGOYA and Tadashi MIYAZAKI  \\
        {\it Department of Physics, Science University of Tokyo,} \\
        {\it Kagurazaka, Shinjuku-ku, Tokyo 162-8601, Japan}
}

\date{}

\maketitle
\vspace{20mm}

\begin{abstract}
A general action is proposed for the fields of $q$-dimensional differential 
form over the compact Riemannian manifold of arbitrary dimensions. 
Mathematical tools are based on the well-known de Rham-Kodaira 
decomposing theorem on harmonic integral. A field-theoretic action for 
strings, $p$-branes and high-spin fields is naturally derived. 
We also have, naturally, the generalized Maxwell equations 
with an electromagnetic and monopole current on a curved space-time. 
A new type of gauge transformations ({\it dual} gauge transformations) 
plays an essential role for coboundary $q$-forms. \\

\noindent
PACS number(s): 02.40.Ky, 03.50.De, 11.10.Kk
\end{abstract}
\setlength{\baselineskip}{7.8mm}
\newpage

\section{Introduction}
\hspace*{6mm}
It goes without saying that field theories play a central role in drawing a 
particle picture. They are especially important to explore a way to construct
a theoretical view on a $curved$ space-time (of more than four dimensions).
Recently-developed theories of strings\cite{1} and membranes\cite{2}, as well 
as those of two-dimensional gravity\cite{3}, go along this way. 
If one makes a complete picture with a general action, one may 
have a clear understanding about why the fundamental structure is ether of 
one dimension (a string), excluding other extended structures
of two or more dimensions, or of other definite dimensions.

The first purpose of this paper is to obtain a general action for
the fields of $q$-dimensional differential forms ({\it $q$-forms}) on a general
curved space-time. In such a way can we deal not only with strings 
and $p$-branes ($p$-dimensional extended objects), 
but also with vector and tensor fields as assigned over each point of a
compact Riemannian manifold (e.g., a sphere or a torus of general dimensions).
 
Our next aim is, as a result of this treatment, to generalize the
conventional Maxwell theory to that on the $curved$ space-time
of {\it arbitrary dimensions}.
Our method is based on the mathematical theory having been developed by de 
Rham and Kodaira\cite{4}. In the theory of harmonic integrals the elegant 
theorem, having been now crowned with the names of the two brilliant 
mathematicians, says that an arbitrary differential form consists of 
three parts: a harmonic form, a $d$-boundary and a $\delta$-boundary. 
With this theorem we have an electromagnetic field coming from the $d$-boundary,
whereas a magnetic monopole field from the $\delta$-boundary. 
We are thus to have a generalized Maxwell theory with 
an electric charge and a magnetic monopole on an arbitrary-dimensional 
curved space-time. Assigning a $\delta$-boundary to a point on the curved 
space-time, we have a new kind of gauge freedom due to the nilpotency 
of the coboundary operator. 

Lastly, we comment on the possibility of 
the case where there could simultaneously exist a matter field and 
a new gauge field interacting together and invariantly under the 
afore-mentioned new type of gauge transformations.

In this paper we proceed to construct a field theory by taking various 
concrete examples. Section 2 treats an algebraic method for obtaining 
a general action. Sections 3 to 6 are devoted to concrete examples. 
In Sect.7 we comment on the case of interacting matter fields with 
a new gauge field. Two often-used mathematical formulas are listed in 
Appendix.

We hope the method developed here will become one of the steps which one 
makes forward to construct the field theory of all extended objects 
\,\,\,-----\,\,\, strings and $p$\,-branes with or without 
{\it spin degrees of freedom} \,\,\,-----\,\,\, based on algebraic geometry.
\vspace{10mm}
\setcounter{equation}{0}
\setcounter{section}{1}
\section{A general action with $q$-forms}
\hspace*{6mm}
Let us start with a Riemannian manifold $M^n$, where we, observers, live,
and with a submanifold $\bar{M}^m$, where particles live ($n,m$ : dimensions 
of the manifolds; $n\geq m$). Both $M^n$ and $\bar{M}^m$ are
supposed to be $compact$
\,\,\,-----\,\,\, compact only because mathematicians construct
a beautiful theory of
harmonic forms over $compact$ spaces, and de Rham-Kodaira's theorem or Hodge's
theorem has not yet been proven with respect to the differential forms over
{\it non-compact} spaces.

We will admit the space $\bar{M}^m$ of a particle to be a submanifold of $M^n$. For
instance, $\bar{M}^m$ may be a circle or a sphere within an $n$-dimensional
(compact) space $M^n$. The local coordinate systems of $M^n$ and $\bar{M}^m$
shall be denoted by $(x^{\mu})$ and $(u^i)$, respectively 
[$\mu =1,2,...,n; i=1,2,...,m$]\cite{5}.  A point ( $u^1$,$u^2$,...,$u^m$) of 
$\bar{M}^m$ is, at the same time, a point of $M^n$,
so that it is also expressed
by $x^\mu=x^\mu(u^i)$. 
In a conventional quantum field theory, point particles, scalar fields, vector
or higher-rank tensor fields, or spinor fields are attributed to each point
of $\bar{M}^m$. In this view we are to assign a $q$-dimensional differential
form ({\it q-form}) $F^{(q)}$ to each point of $\bar{M}^m$, which is expressed
, as mentioned above, by the local coordinate $(u^1,u^2,...,u^m)$ or by 
$x^{\mu}=x^{\mu}(u^i)$. 
Physical objects \,\,\,-----\,\,\, point particles, strings or
electromagnetic fields \,\,\,-----\,\,\,
should be identified with these $q$-forms.

We then make an action with $F^{(q)}$. One of the candidates for the action $S$
is $(F^{(q)} ,F^{(q)}) \equiv \int_{\bar{M}^m}F^{(q)}*F^{(q)}$, where $*$
means Hodge's star operator transforming a $q$-form into 
an $(m-q)$-form. Expressed with respect to
an {\it orthonormal basis} $\omega _1,\omega _2,...,\omega _m$, 
it is defined by the relation
\begin{equation}
*(\omega _{i_1} \wedge \omega _{i_2} \wedge ... \wedge \omega _{i_q})
= \frac{1}{(m-q)!} \left( \begin{array}{cccccc}
1 & 2 & ... & ... & ... & m \\ i_1 & ... &  i_q  &  j_1 & ... & j_{m-q}
\end{array}
\right) 
\omega _{j_1} \wedge \omega _{j_2} \wedge ... \wedge \omega _{j_{m-q}}, 
\label{2:1}
\end{equation}
where $(....)$ denotes the signature $(\pm)$ of the permutation and
the summation convention over repeated indices is, here and hereafter, always
implied. The inner product $(F^{(q)},F^{(q)})$ is a scalar and shares a 
property of scalarity with the action $S$.
Let us, therefore, admit the action $S$ to be proportional to 
$(F^{(q)},F^{(q)})$ and investigate each case that we confront with in the
conventional theoretical physics. 
Thus we put
\begin{eqnarray}
S &=& (F^{(q)},F^{(q)}) = \int_{\bar{M}^m}\hspace{-2mm}{\cal S} \nonumber \\
&=& \int_{\bar{M}^m}\hspace{-2mm}{\cal L}\hspace{1mm}
du^1\wedge du^2\wedge ...\wedge du^m,
\label{2:2} \\
{\cal S} &\equiv& F^{(q)}\!*F^{(q)} = {\cal L}\hspace{1mm}
du^1 \wedge du^2 \wedge  ... \wedge du^m.
\nonumber
\end{eqnarray}
Here ${\cal S}$ is an {\it action form}, but we will sometimes call it by the 
same name $action$. ${\cal L}$ is interpreted as a Lagrangian density.

According to the well-known de Rham-Kodaira theorem, an arbitrary $q$-form
decomposes into three mutually orthogonal $q$-forms:
\begin{eqnarray}
F^{(q)}=F_{\rm I}^{(q)} + F_{\rm II}^{(q)} + F_{\rm III}^{(q)} ,
\label{2:3}
\end{eqnarray}
where $F_{\rm I}^{(q)}$ is a harmonic form, meaning\cite{6}
\begin{eqnarray}
dF_{\rm I}^{(q)} = \delta F_{\rm I}^{(q)} = 0,
\label{2:4}
\end{eqnarray}
and $F_{\rm II}^{(q)}$ is a $d$-boundary, and $F_{\rm III}^{(q)}$ is a 
$\delta$-boundary (coboundary). Here $\delta$ is Hodge's adjoint operator,
which implies $\delta = (-1)^{m(q-1)+1}*d*$ when operated to $q$-forms over 
the $m$-dimensional space. There exist, therefore, a $(q-1)$-form 
$A_{\rm II}^{(q-1)}$ and a $(q+1)$-form $A_{\rm III}^{(q+1)}$, such that
\begin{eqnarray}
F_{\rm II}^{(q)} = dA_{\rm II}^{(q-1)} \: ; \: F_{\rm III}^{(q)} = 
\delta A_{\rm III}^{(q+1)}.
\label{2:5}
\end{eqnarray}
The action $S$ is (proportional to) $(F^{(q)},F^{(q)})$ ;
\begin{eqnarray}
S &\equiv& (F^{(q)},F^{(q)}) \nonumber \\
&=& (F_{\rm I}^{(q)},F_{\rm I}^{(q)}) + (A_{\rm II}^{(q-1)},
\delta dA_{\rm II}^{(q-1)}) + (A_{\rm III}^{(q+1)},d\delta A_{\rm III}^{(q+1)}),
\label{2:6} \\
S &\equiv& \int _{\bar{M}^m}\hspace{-2mm}{\cal S} = \int\!{\cal L}
\hspace{1mm}du^1 \wedge du^2 \wedge .... \wedge du^m. \nonumber
\end{eqnarray}
The physical meaning of Eq.(\ref {2:6}) is whatever we want to discuss in
this paper and
will be described in detail from now on.
\vspace{10mm}

\setcounter{equation}{0}
\section{Point particles, strings and $p$-branes}

\hspace*{6mm}
We first assign $F^{(0)}=1$ to a point $(u^1,...,u^m)$ of the submanifold 
$\bar{M}^m$, and we always make use of the relative (induced) metric 
$\bar{g}_{ij}$ for $\bar{M}^m$ (so that the intrinsic metric of the 
submanifold is irrelevant).
\begin{eqnarray}
\bar{g}_{ij} \equiv \frac{\partial x^\mu(u)}{\partial u^i}
\frac{\partial x^\nu(u)}{\partial u^j} g_{\mu\nu},
\label{3:1}
\end{eqnarray}
where  $g_{\mu\nu}$ is a metric of the Riemannian space $M^n$.
Since the volume element
$dV \equiv \omega_1\wedge \omega_2\wedge...\wedge \omega_m$ is 
expressed, with respect to the local coordinate $(u^i)$, as
\begin{eqnarray}
dV = \sqrt{\bar{g}}\hspace{1mm}du^1\wedge du^2\wedge...\wedge du^m = *1,
\label{3:2}
\end{eqnarray}
we immediately find
\begin{eqnarray}
(F^{(0)},F^{(0)})= \int_{\bar{M}^m}\hspace{-2mm}\sqrt{\bar{g}}
\hspace{1mm}du^1\wedge du^2\wedge...\wedge du^m,
\label{3:3}
\end{eqnarray}
with $\bar{g}=\det (\bar{g}_{ij})$.

When $n=4$ and $m=1$, we have
\begin{eqnarray}
\bar{g} = g_{\mu\nu} \frac{dx^\mu}{du^1} \frac{dx^\nu}{du^1} = g_{\mu\nu} \dot
{x}^\mu \dot{x}^\nu, 
\label{3:4}
\end{eqnarray}
( $\cdot$ means $d/du^1$), hence
\begin{eqnarray}
(F,F) &=&\int_{\bar{M}^1}\hspace{-2mm}ds , 
\label{3:5} \\
ds^2 &=&  g_{\mu\nu} \dot{x}^\mu \dot{x}^\nu {(du^1)}^2 = g_{\mu\nu} dx^{\mu} 
dx^{\nu} , \nonumber
\end{eqnarray}
which indicates that $(F,F)$ is a conventional action (up to a constant) for a point particle in
a 4-dimensional curved space, with $u^1$, interpreted as a proper time.

On the contrary, if we treat a submanifold $\bar{M}^2$, Eq.(\ref{3:3})
becomes
\begin{eqnarray}
(F^{(0)}, F^{(0)}) = \int _{\bar{M}^2}\hspace{-2mm}\sqrt{\bar{g}}
\hspace{1mm}du^1 \wedge du^2\,,
\label{3:6}
\end{eqnarray}
with
\begin{eqnarray}
\bar{g} = \det ( \frac{\partial x^{\mu}}{\partial u^i} \frac{\partial x^{\nu}}
{\partial u^j} g_{\mu\nu}),
\label{3:7}
\end{eqnarray}
which is just the Nambu-Goto action in a $curved$ space (with $u^1=\tau$ and $
u^2=\sigma$ in the conventional notation).
There, and here, the determinant $\bar{g}$ of an induced metric plays an
essential role.
If we confront with an arbitrary submanifold $\bar{M}^{p+1}$
($p$ : an arbitrary integer $\leq n-1$),
we are to have a $p$-brane, whose action is nothing but
that given by Eq.(\ref{3:3}) with $m=p+1$.

Let us discuss the transformation property of the action or Lagrangian 
density.
The transformation of $\bar{M}^m$ into $\bar{M'}^m$ without changing 
$M^n$\cite{7} means reparametrization.
\begin{eqnarray}
u^i &\rightarrow& u'^i , \nonumber \\
x^{\mu}(u^i) &\rightarrow& x'^{\mu}(u'^i) = x^{\mu}(u^i). \label{3:8}
\end{eqnarray}
By this the volume element Eq.(\ref{3:2}) does not change, so that our
Lagrangian (density) for the $p$-brane is trivially invariant under the 
reparametrization.
If we convert $M^n$ into $M'^n$ without changing $\bar{M}^m$, a general
coordinate transformation
\begin{eqnarray}
x^{\mu}(u^i) \rightarrow x'^{\mu}(u^i)  
\label{3:9}
\end{eqnarray}
is induced, under which $\bar{g}_{ij}$ does not change, because of the 
transformation property of the metric $g_{\mu\nu}$.
Our action is trivially invariant also for this general coordinate 
transformation.

If we transform $\bar{M}^m$ and $M^n$ simultaneously, i.e.,
\begin{eqnarray}
u^i &\rightarrow& u'^i , \nonumber \\
x^{\mu}(u^i) &\rightarrow& x'^{\mu}(u'^i),
\label{3:10}
\end{eqnarray}
we do not have an equality $x'^{\mu}(u'^i) = x^{\mu}(u^i)$.
This type of transformations is examined, as an example, for $n=3$ and $m=2$
as follows. Let us take $M^3={\sf R}^3$ (compactified), and
$\bar{M}^2={\sf S}^2$ (2-dimensional surface of a sphere)
whose local coordinate system
is $(u^1 , u^2)$. A point of ${\sf S}^2$ is expressed by $(u^1 , u^2)$, but
it is at the same time a point $(x^1 , x^2, x^3)$ of ${\sf R}^3$. 
We give the relation between the two coordinate systems by the
stereographic projection:
\begin{eqnarray}
x^1 &=& \frac{2r^2 u^1}{(u^1)^2 + (u^2)^2 + r^2}\,, \nonumber \\
x^2 &=& \frac{2r^2 u^2}{(u^1)^2 + (u^2)^2 + r^2}\,, \\
x^3 &=& \frac{r[r^2-(u^1)^2 - (u^2)^2]}{(u^1)^2 + (u^2)^2 + r^2}\,, \nonumber 
\label{3:11}
\end{eqnarray}
where $r$ is the radius of the sphere defining ${\sf S}^2$. The transformation
$(u^1, u^2) \rightarrow (u'^1,u'^2)$ induces the transformation 
$(x^1, x^2, x^3) \rightarrow (x'^1, x'^2, x'^3)$, and vice versa.
The definition of the metric $g_{\mu\nu}$ for $M^n$ and the induced one 
$\bar{g}_{ij}$ for $\bar{M}^m$ tells us 
\begin{eqnarray}
g'_{\mu\nu}(x') = \frac{\partial x^{\rho}}{\partial x'^{\mu}} \frac{\partial x
^{\delta}}{\partial x'^{\nu}} g_{\rho\delta} (x) , 
\label{3:12}
\end{eqnarray}
and hence
\begin{eqnarray}
\bar{g}'_{ij}(u') = \frac{\partial u^k}{\partial u'^i} \frac{\partial u^l}
{\partial u'^j} \bar{g}_{kl} (u) , 
\label{3:13}
\end{eqnarray}
so that we have
\begin{eqnarray}
\sqrt{\bar{g}'(u')}\hspace{1mm}du'^1 \wedge ... \wedge du'^m = \sqrt{\bar{g}(u
)} \hspace{1mm}du^1 \wedge ... \wedge du^m ,
\label{3:14}
\end{eqnarray}
meaning the invariance of the action.
\vspace{10mm}

\setcounter{equation}{0}
\section{Scalar fields}

\hspace*{6mm}
Now we consider the case where a scalar field $\phi (x^{\mu}(u^i))$ is 
assigned to each point $x^{\mu}(u^i)$. From now on we regard every 
quantity as that given over the subspace $\bar{M}^m$, hence we will write
the field simply as $\phi (u^i)$ or $\phi (u)$ instead of 
$\phi (x^{\mu}(u^i))$, etc.

An arbitrary 0-form \,\,\,-----\,\,\, a scalar field \,\,\,-----\,\,\,
decomposes into two parts: 
\begin{eqnarray}
F^{(0)} = F_{\rm I}^{(0)} + F_{\rm III}^{(0)}.
\label{4:1}
\end{eqnarray}
$F_{\rm I}^{(0)}$ is given by
\begin{eqnarray}
F_{\rm I}^{(0)} = \phi (u) ,
\label{4:2}
\end{eqnarray}
with which we obtain 
\begin{eqnarray}
(F_{\rm I}^{(0)} , F_{\rm I}^{(0)}) = \phi ^2(u) dV ,  
\label{4:3}
\end{eqnarray}
meaning a mass term of a scalar field. $F_{\rm III}^{(0)}$ is composed, on the
contrary, of a $\delta$ -boundary of a 1-form:
\begin{eqnarray}
F_{\rm III}^{(0)} &=& \delta A^{(1)} , \nonumber \\
A^{(1)} &=& A_i du^i .
\label{4:4}
\end{eqnarray}
Hence we have
\begin{eqnarray}
F_{\rm III}^{(0)} = -\partial _k (\sqrt{\bar{g}} A^k) \sqrt{\bar{g}} \bar{g} ^
{11} \bar{g}^{22} ... \bar{g}^{mm} ,
\label{4:5}
\end{eqnarray}
where, as usual,
\begin{eqnarray}
A^k = \bar{g}^{kl} A_l \:\: {\rm and} \:\: \partial _k = \frac{\partial}
{\partial u^k} ,
\label{4:6}
\end{eqnarray}
and $(\bar{g} ^{ij})$ is the inverse of $(\bar{g} _{ij})$. In a special
case, where we work out with a flat space and an orthonormal basis, i.e.,
\begin{eqnarray}
\bar{g} ^{ij} = \delta ^{ij} \:\: {\rm and} \:\: du^i = \omega ^i , 
\label{4:7}
\end{eqnarray}
we have a simple form
\begin{eqnarray}
F_{\rm III} ^{(0)} = - \partial _k A^k  ,  
\label{4:8}
\end{eqnarray}
by which the action form ${\cal S}$ becomes
\begin{eqnarray}
{\cal S} = (\partial _k A^k)^2 dV . 
\label{4:9}
\end{eqnarray}
This is the `kinetic' term of the {\it k-vector field} $A^k$.

The {\it gauge transformation} exists for this field: 
\begin{eqnarray}
A^{(1)} &\rightarrow& \tilde{A}^{(1)} = A^{(1)}+\delta A^{(2)} , \nonumber \\
A^{(2)} &=& \frac{1}{2} A_ {{i_1}{i_2}}du^{i_1}\wedge du^{{i_2}} .
\label{4:10}
\end{eqnarray}
In components, it is written as
\begin{eqnarray}
\tilde{A} _h\! =\! A_h\!\! +\! \frac{1}{2(m-2)!} \left( \begin{array}{ccccc}
h & l_1 & ... & ... & l_{m-1} \\
i_1 & i_2 & j_1 & ... & j_{m-2} 
\end{array}
\right )\!
\frac{\partial (\sqrt{\bar{g}} A^{i_1 i_2})}{\partial u^k} 
\sqrt{\bar{g}}\bar{g}^{k l_1}\bar{g}^{l_2 j_1}\!...
\bar{g}^{l_{m-1} j_{m-2}} . 
\label{4:11}
\end{eqnarray}
One can further calculate, if one wants to, to have a beautiful form:
\begin{eqnarray}
\tilde{A} _i &=& A_i - \frac{1}{2} \left( \begin{array}{cc}
j_1 & j_2 \\
k & i 
\end{array}
\right )
\bar{g} ^{kl} D_l A_{j_1 j_2}, \nonumber \\
D_l A_{j_1 j_2} &=& \frac{\partial A_{{j_1}{j_2}}}{\partial u^l} - A_{k j_2} 
\Gamma_{{j_1}l}^k-A_{{j_1}k} \Gamma_{{j_2}l}^k , 
\label{4:12}
\end{eqnarray}
where $\Gamma_{jk}^i$ is the well-known affine connection.
\begin{eqnarray}
\Gamma_{jk}^i = \frac{1}{2} \bar{g}^{il} (\frac{\partial \bar{g} _{jl}}{\partial u^k
} + \frac{\partial \bar{g} _{lk}}{\partial u^j} - \frac{\partial \bar{g} _{jk}
}{\partial u^l} ) .
\label{4:13}
\end{eqnarray}
Note that our fundamental fields are the $A_i$, and the gauge transformation 
is obtained with the $A_{i_1 i_2}$ of the rank {\it higher by one} than the 
former. This is, of course, due to the nilpotency of $\delta$, $\delta ^2 = 0$,
and typical of our new type of formulation. Let us call, here and hereafter, 
that new kind of gauge transformations {\it dual gauge transformations}.
\vspace{10mm}

\setcounter{equation}{0}
\section{Vector fields}

\hspace*{6mm}
When a 1-form $F^{(1)}$ is assigned to each point of $\bar{M}^m$, we have
\begin{eqnarray}
F^{(1)} = F_{\rm I}^{(1)} + F_{\rm II}^{(1)} + F_{\rm III}^{(1)} .
\label{5:1}
\end{eqnarray}
First we will see the contribution of $F_{\rm I}^{(1)}$ to the action,
which is harmonic. Writing as 
\begin{eqnarray}
F_{\rm I}^{(1)} = F_i du^i ,
\label{5:2}
\end{eqnarray}
we immediately have an action (form)
\begin{eqnarray}
{\cal S}_{\rm I} = F_{\rm I}^{(1)}*F_{\rm I}^{(1)}=F_iF^i\!
\sqrt{\bar{g}}\hspace{1mm}
du^1 \wedge ... \wedge du^m 
\label{5:3}
\end{eqnarray}

The contribution of the $d$-boundary is calculated in the same way. Putting
\begin{eqnarray}
F_{\rm II}^{(1)} = dA^{(0)} , 
\label{5:4}
\end{eqnarray}
we have the action
\begin{eqnarray}
{\cal S}_{\rm II} = \bar{g}^{ij} \partial _i A^{(0)} 
\partial _j A^{(0)}\!\sqrt{\bar{g}}\hspace{1mm}du^1 \wedge ... \wedge du^m , 
\label{5:5}
\end{eqnarray}
which expresses a massless scalar particle $A^{(0)}$. Freedom of the choice of
gauges does not here appear. 

The contribution of the $\delta$-boundary is, on the contrary, rather
complicated in calculation. If we put
\begin{eqnarray}
F_{\rm III} ^{(1)} &=& \delta A^{(2)} , \nonumber \\
A^{(2)} &=& \frac{1}{2} A_{i_1 i_2} du^{i_1} \wedge du^{i_2} ,
\label{5:6} \\
F_{\rm III} ^{(1)} &=& F_i du^i , \nonumber 
\end{eqnarray}
we have
\begin{eqnarray}
F_h &=& \left( \begin{array}{ccccc}
h & l_1 & l_2 & ... & l_{m-1} \\
i_1 & i_2 & j_1 & ... & j_{m-2}
\end{array}
\right) 
\frac {\partial}{\partial u^k} ( \sqrt{\bar{g}}A^{i_1 i_2})\sqrt{\bar{g}} 
\,\bar{g}^{kl_1} \bar{g}^{j_1 l_2} ... \bar{g}^{j_{m-2} l_{m-1}} \nonumber \\
&=& - \frac{1}{2} \left( \begin{array}{cc} j_1 & j_2 \\ k & h 
\end{array} \right)
\bar{g}^{kl} D_l A_{{j_1}{j_2}} , 
\label{5:7}
\end{eqnarray}
with $D_l$, defined in Eq.(\ref{4:12})\cite{8}. 
The action is
\begin{eqnarray}
{\cal S}_{\rm III} = F_i F^i\!\sqrt{\bar{g}}\hspace{1mm}
du^1 \wedge ... \wedge du^m . 
\label{5:8}
\end{eqnarray}

The {\it dual} gauge transformation is given in this case by 
\begin{eqnarray}
A^{(2)} &\rightarrow& \tilde{A}^{(2)} = A^{(2)} + \delta A^{(3)} ,
\nonumber \\
A^{(3)} &=& \frac{1}{3!} A_{i_1 i_2 i_3}du^{i_1} \wedge du^{i_2}
\wedge du^{i_3} ,
\label{5:9}
\end{eqnarray}
which trivially leads to the relation
\begin{eqnarray} 
F^{(1)}_{\rm III} = \delta A^{(2)} = \delta \tilde{A}^{(2)} . 
\label{5:10}
\end{eqnarray}
When expressed in components, it is written as
\begin{eqnarray}
\tilde{A} _{h_1 h_2} = 
A_{h_1 h_2} \!\!\!&-&\!\!\! \frac{1}{3!(m-3)!}
\left(
\begin{array}{cccccc}
i_1 & i_2 & i_3 & j_1 & ... & j_{m-3} \\
h_1 & h_2 & l_1 & ... & ... & l_{m-2} 
\end{array}
\right)
\frac{\partial}{\partial u^k} ( \sqrt{\bar{g}} A^{i_1 i_2 i_3}) \nonumber \\
&\times&\!\!\!\sqrt{\bar{g}}\,\bar{g}^{kl_1}\bar{g}^{j_1 l_2}...\bar{g}^{j_{m-
3} l_{m-2}}, 
\label{5:11}
\end{eqnarray}
where, of course, the components with superscript are related to those with
subscript in a conventional manner, as has been described repeatedly.
\begin{eqnarray}
A^{i_1 i_2 i_3} = \bar{g}^{i_1 j_1} \bar{g}^{i_2 j_2} \bar{g}^{i_3 j_3} A_{j_1
 j_2 j_3} .
\label{5:12}
\end{eqnarray}
We finally express Eq.(\ref{5:11}) in an elegant form.
\begin{eqnarray}
\tilde{A} _{h_1 h_2} &=& A _{h_1 h_2} - \frac{1}{3!} 
\left(
\begin{array}{ccc}
j_1 & j_2 & j_3 \\
k & h_1 & h_2
\end{array}
\right)
\bar{g}^{kl} D_l A_{j_1 j_2 j_3} , \nonumber \\
D_l A_{j_1 j_2 j_3} &=& 
\frac{\partial A_{j_1 j_2 j_3}}{\partial u^l} - A_{kj_2 j_3} \Gamma _{j_1 l}^k
- A_{j_1 k j_3} \Gamma _{j_2 l}^k - A_{j_1 j_2 k} \Gamma _{j_3 l} ^k. 
\label{5:13}
\end{eqnarray}

Especially when the space-time is flat and one takes an orthonormal reference
frame, one has 
\begin{eqnarray}
F_i = -\frac{1}{2} \left( \begin{array}{cc}
k & i \\
i_1 & i_2
\end{array}
\right)
\frac{\partial A^{i_1 i_2}}{\partial u^k} ,
\label{5:14}
\end{eqnarray}
which further reduces to a familiar form for $m=4$: 
\begin{eqnarray}
F^i &=& \partial _k A^{ik} , \nonumber \\
{\cal S} &=& \partial _k A^{ik} \partial ^l A_{il} dV.
\label{5:15}
\end{eqnarray}
The {\it dual} gauge transformation becomes in this case
\begin{eqnarray}
\tilde{A} _{i_1 i_2} = A_{i_1 i_2} - \partial _k A_{i_1 i_2 k}  .
\label{5:16}
\end{eqnarray}

Needless to say, the total action comes from adding ${\cal S}_{\rm I}$,${\cal 
S}_{\rm II}$ and ${\cal S}_{\rm III}$. 
A new type of gauge transformations Eq.(\ref{5:13}) appears, due to the
coboundary property of $F_{\rm III} ^{(1)}$.
\vspace{10mm}

\setcounter{equation}{0}
\section{Tensor fields}

\hspace*{6mm}
Now we come to the case where a 2-form is assigned to each point of 
$\bar{M}^m$, the case of which is most useful and attractive for future
development. 

A 2-form decomposes, as usual, into the following three:
\begin{eqnarray}
F^{(2)} = F_{\rm I} ^{(2)} + F_{\rm II} ^{(2)} + F_{\rm III} ^{(2)}\,.
\label{6:1}
\end{eqnarray}
The harmonic form $F_{\rm I} ^{(2)}$ is written with the components
$A_{ij}$ as follows:
\begin{eqnarray}
F_{\rm I} ^{(2)} = \frac{1}{2} A_{i_1 i_2} du^{i_1} \wedge du^{i_2} ,
\label{6:2}
\end{eqnarray}
from which we have
\begin{eqnarray}
{\cal S}_I = F_{\rm I} ^{(2)} * F_{\rm I} ^{(2)} = 
\frac{1}{2} A_{i_1 i_2} A^{i_1 i_2}\!\sqrt{\bar{g}}\hspace{1mm}
du^1 \wedge ... \wedge du^m . 
\label{6:3}
\end{eqnarray}

The contribution of the $d$ -boundary is expressed with our fundamental 1-form
$A^{(1)}$. 
\begin{eqnarray}
F_{\rm II} ^{(2)} = dA ^{(1)} .
\label{6:4}
\end{eqnarray}
This further reduces, when written in components, 
\begin{eqnarray}
F_{\rm II} ^{(2)} &=& \frac{1}{2} F_{i_1 i_2} du^{i_1} \wedge du^{i_2} ,
\nonumber\\
A^{(1)} &=& A_i du^i ,
\label{6:5}
\end{eqnarray}
to a familiar relation
\begin{eqnarray}
F_{ij} = \partial _i A_j - \partial _j A_i , 
\label{6:6}
\end{eqnarray}
which shows that $F_{ij}$
is a field-strength.
The gauge transformation here is given by
\begin{eqnarray}
A^{(1)} \rightarrow \tilde{A} ^{(1)} = A^{(1)} + dA^{(0)} . 
\label{6:7}
\end{eqnarray}
Namely, it is expressed in components as
\begin{eqnarray}
\tilde{A}_i = A_i + \partial _i A(u), 
\label{6:8}
\end{eqnarray}
with  $A(u)$, an arbitrary scalar function, which is a familiar form in the
conventional Maxwell electromagnetic theory. The invariance of the
contribution to $F_{\rm II} ^{(2)}$ owes self-evidently, to the
nilpotency $d^2=0$. 

Let us now add a source term $-2(A^{(1)} , J^{(1)})$ to the action 
with $J^{(1)}$, a source of one -form. Then we have the equation of 
motion from Hamilton's principle of least action:  
\begin{eqnarray}
\delta F_{\rm II} ^{(2)} = \delta dA^{(1)} = J^{(1)}.
\label{6:9}
\end{eqnarray}
In component it is written as follows:
\begin{equation}
- \frac{1}{2} \frac{1}{(m-2)!} \left( \begin{array}{ccccc}
h & l_1 & l_2 & ... & l_{m-1} \\
i_1& i_2& j_1& ...& j_{m-2} \end{array} \right)
\frac{\partial}{\partial u^k} ( \sqrt{\bar{g}} F^{i_1 i_2}) \sqrt{\bar{g}} 
\hspace{1mm}\bar{g} ^{l_1 k} \bar{g} ^{l_2 j_1} ...
\bar{g} ^{l_{m-1} j_{m-2}} = J_h .
\label{6:10}
\end{equation}
After some lengthy calculations we finally have the following beautiful
form. 
\begin{eqnarray}
- \frac{1}{2} \left( \begin{array}{cc} i_1 & i_2 \\ j & h \end{array} 
\right) \bar{g}^{jl} D_l F_{i_1 i_2} = J_h  .
\label{6:11}
\end{eqnarray}

The covariant derivative $D_l$ is given in Eq.(\ref{4:12}).
Equation (\ref{6:10})
or (\ref{6:11}) becomes simple for the $flat$ $m$-dimensional space,
expressed 
in an {\it orthonormal basis}.
\begin{eqnarray}
F_{ij},^j = J_i
\label{6:12}
\end{eqnarray}
This is nothing but the Maxwell equation in an $m$-dimensional space,
with $J_i$, interpreted as an electromagnetic current density. One 
therefore finds that Eq.(\ref{6:9}) or (\ref{6:11}) is the generalized
Maxwell equation in the {\it curved m-dimensional space}.

Here we note that the invariance of the action under the gauge 
transformtion (\ref{6:7}) or (\ref{6:8}) is evident as long as the eqation
for the current
\begin{eqnarray}
\delta J^{(1)} = 0
\label{6:12'}
\end{eqnarray}
holds. In case of the flat space with an orthonormal basis, 
this reduces to the usual form of the conservation of 
current $ \partial _i J^i = 0 $.

Now comes the contribution of the $\delta$-boundary:
\begin{eqnarray}
F_{\rm III} ^{(2)} = \delta A^{(3)},
\label{6:13}
\end{eqnarray}
where $A^{(3)}$ is a 3-form. Expressed, as usual, in components
\begin{eqnarray}
F_{\rm III} ^{(2)} &=& \frac{1}{2} F_{i_1 i_2} du^{i_1} \wedge du^{i_2} ,
\nonumber \\
A^{(3)} &=& \frac{1}{6} A_{i_1 i_2 i_3} du^{i_1} \wedge du^{i_2}
\wedge du^{i_3} ,
\label{6:14}
\end{eqnarray}
Eq.(\ref{6:13}) leads us to 
\begin{eqnarray}
F_{h_1 h_2} \hspace{-2mm}&=&\hspace{-2mm}
-\frac{1}{6(m-3)!} \left( \begin{array}{cccccc}
h_1& h_2& l_1& ...&...& l_{m-2} \\ i_1& i_2& i_3& j_1& ...& j_{m-3}
\end{array} \right) 
\nonumber \\
&&\hspace{20mm}\times
\frac{ \partial}{\partial u^k}(\sqrt{\bar{g}} A^{i_1 i_2 i_3}) 
\sqrt{\bar{g}}\hspace{1mm}
\bar{g}^{l_1 k} \bar{g}^{l_2 j_1}\!...\bar{g}^{l_{m-2} j_{m-3}}.
\label{6:15}
\end{eqnarray}
Along the same line already mentioned repeatedly we further have
\begin{eqnarray}
F_{i_1 i_2} = - \frac{1}{6} \left( \begin{array}{ccc} 
j_1 & j_2 & j_3 \\ k & i_1 & i_2 \end{array} \right) \bar{g}^{kl} D_l
A_{j_1 j_2 j_3}, 
\label{6:16}
\end{eqnarray}
with the covariant derivative $D_l A_{j_1 j_2 j_3}$, defined in 
Eq.(\ref{5:13}).
In the same way as in the case of the $d$-boundary, we add a source term 
$ 2 ( A^{(3)} , *K^{(m-3)} ) $ to the action (\ref{6:3}).
The variation of $A^{(3)}$ gives us the following equation of motion: 
\begin{eqnarray}
dF_{\rm III} ^{(2)} &=& d \delta A^{(3)} = -*K^{(m-3)} , \nonumber \\
K^{(m-3)} &=& \frac{1}{(m-3)!} K_{i_1 i_2 ... i_{m-3}} du^{i_1} \wedge ...
\wedge du^{i_{m-3}} ,
\label{6:17}
\end{eqnarray}
One has the relation between the components of $F_{\rm III} ^{(2)}$ and
$K^{(m-3)}$:
\begin{equation}
F_{i_1 i_2 , i_3} + F_{i_2 i_3 , i_1} + F_{i_3 i_1 , i_2} = 
- \frac{1}{(m-3)!} \left( \begin{array}{cccccc} 
1& 2& ...&...&...&m \\ j_1& ...& j_{m-3}& i_1& i_2& i_3 \end{array} \right)
\hspace{-1mm}\sqrt{\bar{g}} K^{j_1 ... j_{m-3}} ,
\label{6:18}
\end{equation}
where $F_{i_1 i_2 , i_3} \equiv \partial F_{i_1 i_2} / \partial\, u^{i_3}$, etc..
If our space-time $\bar{M}^m$ is {\it flat and the dimension is} $m=4$, these
expressions reduce to a familiar form.
\begin{eqnarray}
F_{\mu \nu} &=& -\partial ^{\rho} A_{\mu \nu \rho} , \nonumber \\
\tilde{F} _{\mu \nu} ,^{\nu} &=& K_{\mu} ,
\label{6:19}
\end{eqnarray}
where
\begin{eqnarray}
\tilde{F} _{\mu \nu} &=& \frac{1}{2} \epsilon _{\mu \nu \rho \sigma} F^{\rho 
\sigma} , \nonumber \\
K^{(1)} &=& K_{\mu} du^{\mu} .
\label{6:20}
\end{eqnarray}
Equations (\ref{6:19}) and (\ref{6:20}) tell us that $K_{\mu}$
is a {\it magnetic monopole current}\cite{10}.

The {\it dual gauge transformation} is, in this case, given by
\begin{eqnarray}
A^{(3)} \rightarrow \tilde{A}^{(3)} = A^{(3)} + \delta A^{(4)} .
\label{6:21}
\end{eqnarray}
In components is it written as
\begin{eqnarray}
\tilde{A}_{h_1 h_2 h_3} \hspace{-2mm}&=&\hspace{-2mm}A_{h_1 h_2 h_3}+
\frac{1}{4!(m-4)!}
\left( \begin{array}{ccccccc}
h_1& h_2& h_3& l_1& ...&...& l_{m-3} \\ i_1& i_2& i_3& i_4& j_1& ...& j_{m-4} 
\end{array} \right) \nonumber \\
&&\times\frac{\partial}{\partial u^k} ( \sqrt{\bar{g}} A^{i_1 i_2 i_3 i_4})
\sqrt{\bar{g}}\hspace{1mm}
\bar{g}^{l_1 k} \bar{g}^{l_2 j_1} ... \bar{g}^{l_{m-3} j_{m-4}},
\label{6:22}
\end{eqnarray}
which one can further rewrite as follows:
\[
\tilde{A}_{i_1 i_2 i_3}=A_{i_1 i_2 i_3} + \frac{1}{4!} 
\left( \begin{array}{cccc}
j_1 & j_2 & j_3 &j_4 \\ k & i_1 & i_2 & i_3 \end{array} \right)
g^{kl} D_l A_{j_1 j_2 j_3 j_4},
\]
\begin{equation}
D_l A_{j_1 j_2 j_3 j_4}=\frac{\partial A_{j_1 j_2 j_3 j_4}}{\partial u^l} -  A
_{k j_2 j_3 j_4} \Gamma _{j_1 l}^k - A_{j_1 k j_3 j_4} \Gamma _{j_2 l}^k - A_{
j_1 j_2 k j_4} \Gamma _{j_3 l}^k - A_{j_1 j_2 j_3 k} \Gamma _{j_4 l}^k .
\label{6:23}
\end{equation}

The invariance of the action under the dual gauge transformation (\ref{6:21}) 
is assured for the current $K^{(m-3)}$ that satisfies
\begin{eqnarray}
d K^{(m-3)} = 0. 
\label{6:23'}
\end{eqnarray} 

The action form ${\cal S}_{\rm III} = F_{\rm III}^{(2)} * F_{\rm III}^{(2)}$ 
can be, of course, calculated along the same line already mentioned.
And the total action $S$ is
\begin{eqnarray}
S = S_{\rm I} + S_{\rm II} + S_{\rm III} -2(A^{(1)} , J^{(1)}) 
+ 2(A^{(3)} , *K^{(m-3)})
\label{6:24}
\end{eqnarray}
\vspace{10mm}

\setcounter{equation}{0}
\section{Comments and discussions}

\hspace*{6mm}
We have taken, up to now, the position that we only have a gauge field 
(or a scalar Field) ($q$-form $ F^{(q)}$) as a fundamental field. 
Here the question arises as to whether there can simultaneously exist 
a matter filed and a gauge field at the outset, both of which, 
together, interact with each other gauge-invariantly.
This viewpoint is conventional, but a dual gauge transformation should be 
introduced if one has a well-defined $\delta$-boundary.

Unfortunately, we are led to a very restricted way of treating. As a simple 
example we manipulate an ($m-2$)-form $F^{(m-2)}_{\rm III}$ 
({\it $\delta$}-boundary) 
and a two-component real field $\phi^A (u)$ [$A=1,2$], assigned to each point 
of the manifold $\bar{M}^m$, i.e.,
\begin{eqnarray}
  F^{(m-2)}_{\rm III} & = & \delta A^{(m-1)}, \nonumber \\
  A^{(m-1)}           & = & {\displaystyle \frac{1}{(m-1)!}} A_{i_{1} \cdots i_{m-1}} d{u^{i_{1}}} \wedge \cdots \wedge  d{u^{i_{m-1}}}.
\label{eq:7.1}
\end{eqnarray}
The dual gauge transformation is given by
\begin{equation}
A^{(m-1)} \longrightarrow {\tilde{A}}^{(m-1)}=A^{(m-1)} + {\delta A^{(m)}}. 
\label{eq:7.2}
\end{equation}
In components we have 
\begin{eqnarray}
   {\tilde{A}}_{i_{1} \cdots i_{m-1}}   & = & A_{i_{1} \cdots i_{m-1}} -
                      \left(
                        \begin{array}{cccc}
                                  1 &   2   & \cdots &   m    \\
                                  k & i_{1} & \cdots & i_{m-1}    
                        \end{array}  
                     \right)  {\bar{g}}^{kl} D_{l} A_{1 2 \cdots m}, \nonumber \\
D_{l} A_{1 2 \cdots m} & = & {\displaystyle \frac{{\partial  A_{1 2 \cdots m} }}{\partial u^l}} - A_{1 2 \cdots m} {\Gamma}^{k}_{k l} \nonumber \\
                       & = & {\displaystyle \frac{{\partial  A_{1 2 \cdots m} }}{\partial u^l}} - \frac{1}{2} A_{1 2 \cdots m} \frac{\partial}{\partial u^l} \ln \bar{g}.
\label{eq:7.3}
\end{eqnarray}
With these $A_{i_{1} \cdots i_{m-1}}$ we introduce a {\it dual one-form } 
$B^{(1)}$ whose components are $B_{j}$ as follows.
\begin{eqnarray}
B^{(1)} & \equiv & \ast A^{(m-1)}, \nonumber \\
B_{j}   &    =   & {\displaystyle \frac{1}{(m-1)!}}
              \left(
                \begin{array}{cccc}
                     1    & \cdots &   m-1   & m     \\
                   i_{1}  & \cdots & i_{m-1} & j    
                \end{array}        
              \right)  
             {\sqrt{\bar{g}}} A^{i_{1} \cdots i_{m-1}}. 
\label{eq:7.4}
\end{eqnarray}
The field strength $B_{ij}$ is given by 
\begin{equation}
B_{ij} \equiv {\partial}_{i} B_{j} - {\partial}_{j} B_{i}.  \label{eq:7.5}
\end{equation}
The dual gauge transformation reduces to a conventional form with this dual 
one-form;
\begin{equation}
{\tilde{B}}_{k}=B_{k} + {\partial}_{k} \lambda,  \label{eq:7.6}
\end{equation}
where $\lambda$, a scalar dual to $A^{i_{1} \cdots i_{m}}$ , is 
\begin{eqnarray}
\lambda(u) & \equiv & - {\displaystyle \frac{1}{m!}}
        \left(
          \begin{array}{cccc}
              1    & \cdots &   m-1   &   m     \\
            i_{1}  & \cdots & i_{m-1} & i_{m}    
          \end{array}        
        \right) 
        {\sqrt{\bar{g}}} A^{i_{1} \cdots i_{m-1} i_{m}} \nonumber \\
           &    =   & -\sqrt{\bar{g}} A^{1 2 \cdots m}.
\label{eq:7.7}
\end{eqnarray}
The $\delta$-boundary $F^{(m-2)}_{\rm III}$ is calculated 
with this $B^{(1)}$ to be
\begin{eqnarray}
\lefteqn{ 
F^{(m-2)}_{\rm III} =-\frac{1}{2(m-2)!}
        \left(
          \begin{array}{ccccc}
              1   &   2   &   3    & \cdots &    m     \\
            h_{1} & h_{2} & l_{1}  & \cdots & l_{m-2}
          \end{array}  
        \right)   } \nonumber  \\
    & & \hspace{2cm} \times  {\sqrt{\bar{g}}} B_{kj}  {\bar{g}}^{k h_{1}} {\bar{g}}^{j h_{2}} d{u^{l_{1}}} \wedge \cdots \wedge d{u^{l_{m-2}}}. \label{eq:7.8}
\end{eqnarray}
The local U(1) gauge transformation for the matter field $\phi^{A} (u)$ is 
obtained, 
with $\lambda (u)$ now infinitesimal,
\begin{eqnarray}
{\hat{\delta}} {\phi}^{A} (u)  & = & {\rm T^A}_{B} {\phi}^{B} (u) \lambda (u),   \nonumber \\
         {\rm T}               & = & \left(
                                       \begin{array}{cc}
                                                  0 & 1      \\
                                                 -1 & 0    
                                       \end{array}  
                                     \right).   \label{eq:7.9}
\end{eqnarray} 
Here we have the covariant derivative for $\phi^A (u)$;
\begin{equation}
{\nabla}_{i} {\phi}^{A} (u) = {\partial}_{i} {\phi}^{A} (u) -  {\rm T^A}_{B} {\phi}^{B} (u) B_{i} (u),  \label{eq:7.10}
\end{equation}
and we immediately have the {\it covariance} of ${\nabla}_{i} {\phi}^{A} (u)$;
\begin{equation}
{\hat{\delta}} {\nabla}_{i} {\phi}^{A} (u) = {\rm T^A}_{B} ({\nabla}_{i} {\phi}^{B} (u)) \lambda (u).  \label{eq:7.11}
\end{equation}
The total Lagrangian density is 
\begin{eqnarray}
{\cal{L}}_{\rm tot} & = &  {{\cal{L}}^{(m-2)}_{\rm gauge}} + {\cal{L}}_{\rm matter}, \nonumber \\
{{\cal{L}}^{(m-2)}_{\rm gauge}} & = & - {\displaystyle \frac{1}{2}} {\sqrt{\bar{g}}} {\bar{g}}^{i_{1} j_{1}} {\bar{g}}^{i_{2} j_{2}} B_{i_{1} i_{2}} B_{j_{1} j_{2}}, \nonumber \\
{\cal{L}}_{\rm matter} & = & {\displaystyle \frac{1}{2}} \sqrt{\bar{g}} {\nabla}_{k} {\phi}^{A} {\nabla}^{k} {\phi}_{A} - \frac{1}{2} \sqrt{\bar{g}} {\mu}^2 {\phi}^{A}  {\phi}_{A}, 
\label{eq:7.12}
\end{eqnarray}
with $ \mu $, the mass of the matter field, and we give the Lagrangian for 
the gauge-field sector a minus sign, in order to have a positive energy.

The equations of motion for the matter field $\phi^A (u)$ and the gauge field 
$B_{k}(u)$are, respectively,
\begin{eqnarray}
{\displaystyle {\sqrt{\bar{g}}} B_{k} {\nabla}^{k} {\phi}_{B}  {\rm T^B}_{A} + {\sqrt{\bar{g}}} {\mu}^{2} {\phi}_{A} + {\partial}_{k} ( {\sqrt{\bar{g}}} {\nabla}^{k} {\phi}_{A} )} & = & 0, \nonumber \\
{\displaystyle 2{\partial}_{l} ({\sqrt{\bar{g}}} B^{kl}) - {\sqrt{\bar{g}}} {\nabla}^{k} {\phi}_{A}  {\rm T^A}_{B} {\phi}^{B}} & = & 0. 
\label{eq:7.13}
\end{eqnarray}
It goes without saying that the conserved Noether current exists for our U(1) 
gauge transformation.

One wonders, here, that nothing differs in gauge transformation for $\delta$-
boundary from for the conventional $d$-boundary.
The essential point is that, for and only for $q=m-2$, the dual gauge 
transformation reduces to an ordinary gauge transformation according as 
the $(m-1)$-form $A^{i_{1} \cdots i_{m-1}}$ dually transforms to the vector 
$\sqrt{\bar{g}} B^k$. In this case $F^{(m-2)}_{\rm II}$ ($d$-boundary)
$=d A^{(m-3)}_{\rm II}$, and the gauge transformation of $A^{(m-3)}_{\rm II}$ becomes: 
$A^{(m-3)}_{\rm II} \longrightarrow {\tilde{A}}^{(m-3)}_{\rm II}
=A^{(m-3)}_{\rm II} + {d A^{(m-4)}_{\rm II}}$.
So as this gauge transformation be conventional, we must have $m=4$.
Hence we have the fact that in case of our space-time being $1+3$ dimensional,
we have both electromagnetic and monopole currents as well as 
the matter field.

Now comes the conclusion. The $q$-form formulation over the compact 
Riemannian manifold leads us to the world where both electromagnetic and 
monopole currents exist. The mathematical tool we adopt is based on the
de Rham-Kodaira decomposing theorem of harmonic forms.
Higher-rank $q$-form endows a particle with an intrinsic degree of freedom
(integer sign). In case of $q=m-2$, we are able to introduce both the matter
field and dual gauge field ($\delta$-boundary) from the beginning.
For $m=4$ and $q=2$, we can start with three kinds of fields:
Electromagnetic fields ($d$-boundary), dual fields ($\delta$-boundary) and 
matter fields over the {\it curved }space-time. The last fields are coupled 
with the former two fields; the way of coupling is gauge invariant and 
dual-gauge invariant.
\vspace{10mm}

\section*{Acknowledgement}

\hspace*{6mm}
\noindent
One of the authors (H.E) thanks Iwanami F\=ujukai for financial support.

\newpage
\addcontentsline{toc}{section}{Appendix}
\appendix
\renewcommand{\theequation}{\Alph{section}.\arabic{equation}}
\setcounter{equation}{0}
\section{Hodge's star operator}

\hspace*{6mm}
As defined by Eq.(\ref{2:1}), Hodge's star operator $*$ is an isomorphism of
${\cal H}^q$ (liner space of $q$-forms) into ${\cal H}^{m-q}$ .
Here, in this appendix, we only write down two important formulas which 
we frequently use in calculation in Sects.4 to 7.

For an arbitrary $q$-form
\begin{eqnarray}
\varphi = \frac{1}{q!} \varphi _{i_1 i_2 ... i_q} du^{i_1} \wedge du^{i_2}
\wedge ... \wedge du^{i_q}, 
\label{A:1}
\end{eqnarray}
we have
\begin{eqnarray}
* \varphi = \frac{1}{(m-q)!q!} \left( \begin{array}{cccccc} 
1&2& ...&...&...& m \\ i_1 &... &i_q& j_1& ... &j_{m-q}  \end{array} \right) 
\sqrt{\bar{g}}\hspace{1mm}
\varphi ^{i_1 ... i_q} du^{j_1} \wedge ... \wedge du^{j_{m-q}}, 
\label{A:2}
\end{eqnarray}
where
\begin{eqnarray}
\varphi ^{i_1 ... i_q} = \bar{g} ^{i_1 l_1} ... \bar{g}^{i_q l_q} \varphi _{l_
1  ... l_q} ,
\label{A:3}
\end{eqnarray}
with $\bar{g}_{ij}$, the metric tensor. 

As for a basis of ${\cal H}^q$, we have 
\begin{eqnarray}
*(du^{k_1} \wedge ... \wedge du^{k_q}) = \frac{1}{(m-q)!} \left( \begin
{array}{cccccc} 1 &2& ...&...&...& m \\ i_1&...& i_q& j_1& ...& j_{m-q}
\end{array} \right) \nonumber \\
\times \sqrt{\bar{g}}\hspace{1mm}\bar{g}^{i_1 k_1}...
\bar{g}^{i_q k_q} du^{j_1} \wedge ... \wedge du ^{j_{m-q}}. 
\label{A:4}
\end{eqnarray}
Note that a factor $1/q!$ is removed here in the right-hand side of Eq.(\ref{A:4}).


\newpage
\begin{thebibliography}{99}
\bibitem{1}
        M.B.~Green, J.H.~Schwarz and E.~Witten,
        {\it Superstring Theory I,II}
	(Cambridge Univ. Press, Cambridge, 1987);\\
	L.~Brink and M.~Henneaux,
	{\it Principles of String Theory}
	(Plenum Press, New York, 1988).
\bibitem{2}
	K.~Kikkawa and M.~Yamasaki,
	Prog.~Theor.~Phys.{\bf 76} (1986) 1379;\\
	J.~Hoppe,
	Elem.~Part.~Res.~J. (Kyoto) {\bf 80} (1989) 145;\\
	M.~Yamanobe,
	``{\it P-Branes in the Extended Picture of Elementary Particles}''
	(Ph.D thesis, Science Univ. of Tokyo, 1996);\\
	S.~Ishikawa, Y.~Iwama, T.~Miyazaki and M.~Yamanobe,
	Int.~J.~Mod.~Phys. {\bf A10} (1995) 4671;\\
	S.~Ishikawa, Y.~Iwama, T.~Miyazaki, K.~Yamamoto, M.~Yamanobe and
	R.~Yoshida,
	Prog.~Theor.~Phys. {\bf 96} (1996) 227.
\bibitem{3}
	C.J.~Isham, R.~Penrose and P.W.~Sciama (Editors),
	{\it Quantum Gravity 2 : a Second Oxford Symposium}
	(Clarendon Press, Oxford, 1981);\\
	F.~David,
	``{\it Simplicial Quantum Gravity and Random Lattices}'',
	in {\it Gravitation and Quantizations}
	(Editors : B.~Julia and J.~Zinn-Justin, Les Houches 1992
	Session LVII, pp.679-750, Elsevier Sci. B.V., 1995);\\
	P.~Pi Francesco, P.~Ginsparg and J.~Zinn-Justin,
	Phys.~Rep. {\bf 254} (1995) 1.        
\bibitem{4}
	Y.~Akizuki,
	{\it Harmonic Integral, 2nd Edition}
	(Iwanami, Tokyo, 1972).
\bibitem{5}
	We will also call the local coordinate system by the name of the
	manifold itself.
\bibitem{6}
	Our manifold is assumed to be compact,
	so that harmonicity reduces to Eq.(2.4).
\bibitem{7}
	We are transforming a local coordinate system into another;
	remember the footnote \cite{5}.
\bibitem{8}
	Here, and henceforth, the components of the tensors
	$A_{{i_1}{i_2}...{i_n}}$ are always antisymmetric with
	respect to the exchange of suffices.
\bibitem{9}
	R.P.~Feynman and J.A.~Wheeler,
	Rev.~Mod.~Phys. {\bf 21} (1949) 425;\\
	M.~Kalb and P.~Ramond,
	Phys.~Rev. {\bf D9} (1974) 2273;\\
	M.~Yamanobe, See Ref.\cite{2}.
\bibitem{10}
	P.A.M.~Dirac,
	Proc.~Roy.~Soc. {\bf A133} (1931) 60;
	Phys.~Rev. {\bf 74} (1948) 817.
\end{thebibliography}
\end{document}